\documentclass[aps,prd,preprint,preprintnumbers,showpacs]{revtex4-1}
\usepackage{enumerate}
\usepackage{graphicx}
\usepackage{amsmath}
\usepackage{caption}
\usepackage[section] {placeins}
\usepackage{slashed}
\usepackage[dvips]{color}
\usepackage{soul}

\begin{document}

\title{ Role of vector and pseudoscalar mesons in understanding $1/2^-$ $N^*$ and $\Delta$ resonances}
\author{K.~P.~Khemchandani$^{1,3}$\footnote{kanchan@if.usp.br}}
\author{A.~Mart\'inez~Torres$^{1,3}$\footnote{amartine@if.usp.br}}
\author{ H.~Nagahiro$^{2, 3}$\footnote{nagahiro@rcnp.osaka-u.ac.jp} }
\author{ A.~Hosaka$^3$\footnote{hosaka@rcnp.osaka-u.ac.jp}}
\preprint{}

 \affiliation{
$^1$ Instituto de F\'isica, Universidade de S\~ao Paulo, C.P 66318, 05314-970 S\~ao Paulo, S\~ao Paulo, Brazil.\\
$^2$ Department of Physics, Nara Women's University,  Nara 630-8506, Japan.\\
$^3$ Research Center for Nuclear Physics (RCNP), Mihogaoka 10-1, Ibaraki 567-0047, Japan.
}

\date{\today}

\begin{abstract}
A study of  nonstrange  meson-baryon systems has been made with the idea of  understanding the properties of the low-lying $1/2^-$ $N^*$ and $\Delta$ resonances.  The coupled channels are built by considering the pseudoscalar and vector mesons together with the octet baryons. The formalism is based on obtaining the  interactions from the lowest order chiral Lagrangian when dealing with pseudoscalar mesons and  relying on  the hidden local symmetry in case of the vector mesons. The transition between the two systems is obtained by replacing the photon by a vector meson in the Kroll-Ruderman theorem  for the photoproduction of pseudoscalar mesons. The subtraction constants, required to calculate the loop-function in the scattering equations, are constrained by fitting the available experimental data on some of the reactions with pseudoscalar meson-baryon  final states. As a consequence, we find resonances which can be related to  $N^*(1535)$, $N^*(1650)$ (with a double pole structure),  $N^*(1895)$ and $\Delta(1620)$.  We conclude that these resonances can be, at least partly, interpreted as dynamically
generated resonances and that  the vector mesons play an important role in determining the dynamical origin of the low-lying  $N^*$ and $\Delta$ states.

\end{abstract}

\maketitle

\section{Introduction}
Information on the properties of the  excited states  of the nonstrange baryons (i.e., those made of  quarks $u$ and $d$)  is one of the most sought-after due to the relevance of the same to the nuclear and low energy hadron physics, which can be  accessed at several  experimental facilities existing around the world.  To state a few examples, these resonances  play an important role in understanding $N-N$  interaction  \cite{nn_int1,nn_int2}, in describing cross sections for the reaction with meson-nucleus  final states (for instance, see Refs.~\cite{mosel,michael,mine,karin}), in approaching some fundamental issues, like, existence of multiquark states \cite{bruno}, occurrence of  OZI violating  processes \cite{ozi}, etc.  Such relevance serves as a motivation for the dedicated efforts made by several groups in extracting information related to $N^*$ and $\Delta$ resonances through different approaches: like, partial wave analysis of the relevant data \cite{pwa}, within quark models \cite{quarkmodel}, unitarized dynamical models  (see, for example, Refs.~\cite{juelich,Nakayama,kamano,lutz,oller,oset,threebody,carmen1,carmen2} and those given in these papers), etc. 

The motivation of the present article is in line with the above mentioned works. To be more specific, we investigate if  the meson-baryon dynamics, where pseudoscalar and vector mesons are considered, plays an important role in understanding the properties of isospin 1/2 and 3/2 nonstrange  baryon resonances, especially the ones  with spin parity $1/2^-$. This paper can be considered as a continuation of our previous studies of meson-baryon dynamics \cite{us1,us2,us3}. In Ref.~\cite{us1} we studied the vector meson-baryon (VB) interaction in detail starting from an SU(2) Lagrangian motivated by the gauge invariance of  the hidden local symmetry (which treats vector mesons as gauge bosons) \cite{bando}. We found that such a  gauge invariance of this Lagrangian compels the consideration of a contact interaction  arising from the same Lagrangian. In addition, contrary to the case of the pseudoscalar-baryon (PB) systems, we found that the baryon exchange (in $s$- and $u$-channels) diagram give a  contribution comparable to the one coming from the $t$-channel diagram (which gives dominant contribution in the PB case). We showed that the sum of such diagrams lead to a spin-isospin dependent VB interaction. Further, a generalization to the SU(3) case was made and the results were found to be different to the ones obtained in Ref.~\cite{eovb}, where  the $t$-channel was considered to study VB systems and several spin-degenerate resonances were found to couple strongly to  vector mesons.  However, we must mention that the work in Ref.~\cite{eovb} has further been extended by including one pion loop contribution to the VB systems  and some interesting results have been found \cite{javi}.

Coming back to our works, we further investigated the importance of coupling PB and VB systems with strangeness $-1$ in Ref.~\cite{us2},  keeping in mind the knowledge that the low-lying strange resonances seem to fit better in a meson-baryon molecular picture ~\cite{hyodohosaka}. In Ref.~\cite{us2} we used a simplified VB interaction since we concentrated on  low-lying resonances (on which more information is available). We found that low-lying resonances couple strongly to VB systems, implying large weight of VBB$^*$ vertices (where $B^*$ represents a baryon resonance) in, for instance, photoproduction processes.  This  finding motivated a full coupled PB-VB channel calculation considering detailed VB interaction (as the one used in Ref.~\cite{us1})  in order to explore higher mass $\Lambda$'s and $\Sigma$'s, which resulted in findings related to several hyperon resonances compatible with information available from experimental studies \cite{us3}.

As a continuation of Refs.~\cite{us1,us2,us3}, and following the formalism developed in these works, here we look at the nonstrange meson-baryon systems.  An analysis of meson-baryon scattering  made in Ref.~\cite{hyodohosaka} shows that the resonances generated in the nonstrange PB systems  \cite{inoue} 
do not seem to relate well with dynamically generated states. 
This is intriguing since producing a light meson, like pion, requires about 140 MeV of energy only while the first negative parity $N^*$ is about 500 MeV heavier than the nucleon, which, intuitively, could have an important contribution from the $\pi N$ interaction and thus be related to a meson-baryon molecular state. In addition, the quark model calculations, for example, of  Isgur and Karl \cite{isgur} did not result in a good reproduction of the properties of $N^*(1535)$, indicating towards something missing  in their framework. In Ref.~\cite{zou_npa}, it has been suggested that the wave function of $N^*(1535)$ may have large $s\bar{s}$ component and might require  5-quark contributions. Yet more studies of PB channels have been made in past, which  can reproduce the poles related to $N^*(1535), N^*(1650)$ by solving Bethe-Salpeter equations as integral ones, by keeping the off-shell nature, \cite{juan,bruns} based on  chiral Lagrangian and by considering contributions from next-to-leading-order terms, although such formalisms introduce additional parameters in the model. One could imagine that it might be possible to interpret these additional parameters if one could keep the contributions to the lowest order Lagrangian and add different seeds (intrinsic ones or from other meson-baryon channels) to the formalism.  

In view of this situation, obtaining new information in this sector, within our formalism, could be useful.  We, thus, try to find an answer to the question: does vector meson-baryon dynamics bring new information related to the nature of the low-lying nonstrange resonances, like $N^*(1535), N^*(1650)$?

\section{Meson-baryon interactions and  scattering equations}\label{formalism}
We make a brief discussion of the formalism in this section since more detailed information can be obtained from Refs.~\cite{us1,us2,us3}. The aim of this work is to study meson-baryon systems with total strangeness zero and, as a standard approach, the framework consists of solving the scattering equations in a coupled channel formalism. In the present work, we couple pseudoscalar and vector mesons, which gives nine channels, in the isospin base, with total strangeness zero: $\pi N$, $\eta  N$, $K \Lambda$, $K \Sigma$, $\rho N$, $\omega N$, $\phi N$, $K^* \Lambda$, $K^* \Sigma$. To start with the study of these systems, we need amplitudes for the processes: PB~$\rightarrow$~PB, VB~$\rightarrow$~VB and PB~$\leftrightarrow$~VB. 

The  PB $\leftrightarrow$ VB transition amplitudes are obtained from a Lagrangian (as deduced in Refs.~\cite{us2,us3}) by using the Kroll-Ruderman theorem for the photoproduction of a pion and by introducing the vector meson as the a gauge boson of the hidden local symmetry)
\begin{eqnarray}
\mathcal{L}_{PBVB} = \frac{-i g}{2 f_\pi} \left ( F \langle \bar{B} \gamma_\mu \gamma_5 \left[ \left[ P, V_\mu \right], B \right] \rangle + 
D \langle \bar{B} \gamma_\mu \gamma_5 \left\{ \left[ P, V_\mu \right], B \right\}  \rangle \right), \label{pbvb}
\end{eqnarray}
where the trace $\langle ... \rangle$ has to be calculated in the flavor space and $F = 0.46$, $D=0.8$ such that  $F + D \simeq  g_A = 1.26$ with $g_A$ denoting the axial coupling of the nucleon, and the ratio 
$D/(F+D) \sim 0.63$. The latter ratio is close to the SU(6) quark model value of 0.6 obtained in Ref.~\cite{Yamanishi:2007zza}. 

The Lagrangian in Eq.~(\ref{pbvb}) leads to the amplitude
\begin{eqnarray}
V^{PBVB}_{ij} = i \sqrt{3} \frac{g_{KR}}{2f_\pi} C^{PBVB}_{ij}, \label{Vpbvb}
\end{eqnarray}
where, using the Kawarabayashi-Suzuki-Riazuddin-Fayazuddin relation \cite{ksrf1,ksrf2}, we get 
\begin{equation}
g_{KR} = m_\rho/\left(\sqrt{2} f_\pi \right) \sim 6\label{gkr}
\end{equation}
with the subscript $KR$ on $g$ indicating  the Kroll-Ruderman coupling. To obtain this value of the coupling we have used $f_\pi = 93$~MeV and  the mass of the rho  meson $m_\rho = 770$ MeV. The coefficients
  $C^{PBVB}_{ij}$ in Eq.~(\ref{Vpbvb}) were not obtained in Refs.~\cite{us2,us3} for the non strange meson-baryon systems. We give this information in the present article, in Tables~\ref{kr_iso1/2} and \ref{kr_iso3/2},  for isospin 1/2 and 3/2, respectively.
\begin{table} [h!]
\caption{ $C^{PBVB}_{ij}$  coefficients of the PB $\rightarrow$ VB amplitude (Eq.~(\ref{Vpbvb})) in the isospin 1/2 configuration.} \label{kr_iso1/2}
\begin{ruledtabular}
\begin{tabular}{cccccc}
&$\rho N$&$\omega N$&$\phi N$&$K^* \Lambda$&$K^*\Sigma$\\
\hline
$\pi N$&$-2\left( D + F \right)$&$0$&$0$&$-\frac{1}{2}  \left( D + 3F \right)$&$\frac{1}{2} \left( F - D \right)$\\
$\eta N$&$0$&$0$&$0$&$\frac{1}{2}  \left( D + 3F \right)$&$\frac{3}{2} \left( F - D \right)$\\
$K \Lambda$&$-\frac{1}{2} \left( D + 3F \right)$ & $\frac{1}{2\sqrt{3}}\left( D + 3F \right)$&$-\frac{1}{\sqrt{6}}\left( D + 3F \right)$&$-D$&$D$\\
$K\Sigma$&$\frac{1}{2}  \left( F - D \right)$ &$\frac{\sqrt{3}}{2} \left( F - D \right)$&$\sqrt{\frac{3}{2}} \left( D - F \right)$&$D$&$D - 2F$\\
\end{tabular}
\end{ruledtabular}
\end{table}

We should mention here that, in our formalism, we can couple PB-VB channels in the spin 1/2 configuration only (in s-wave interaction, which is relevant in the present case since we study dynamical generation of resonances).   Thus  we study isospin 1/2 and 3/2 meson-baryon systems with total spin 1/2 (which can generate $1/2^-$ $N^*$'s and $\Delta^*$'s).  
\begin{table} [htbp]
\caption{$C^{PBVB}_{ij}$  coefficients of the PB $\rightarrow$ VB amplitude (Eq.~(\ref{Vpbvb})) in the isospin 3/2 configuration.}\label{kr_iso3/2}
\centering
\begin{ruledtabular}
\begin{tabular}{ccccccc}
&$\rho N$&$K^* \Sigma$\\
\hline
$\pi N$&$\left( D + F \right)$&$\left( F - D \right)$\\
$K \Sigma$&$\left( F - D \right)$&$\left( D + F \right)$\\
\end{tabular}
\end{ruledtabular}
\end{table}

Going over to the discussion of VB interactions, it was shown in Ref.~\cite{us1} that starting with the Lagrangian for the $\rho N$ interaction, which includes the vector and tensor terms
\begin{equation}
\mathcal{L} =  \bar{N} \left( i \slashed \partial - g F_1 \gamma_\mu \rho^\mu \right) N,\label{rhon}
\end{equation}
and which is consistent with  the gauge invariance of hidden local symmetry, one ends with equally important contributions from 
 $s$-, $t$-, and $u$-channel exchange diagrams together with the contact term (CT) arising from the commutator in the vector meson tensor.  Here we should remind the reader that we are considering the exchange of $1/2^+$ octet baryons in s- and u-channel diagrams which, in case of s-wave meson-baryon interaction, gets contribution only from the negative energy solution of the Dirac equation (giving rise to the corresponding ``$Z$-diagrams"). This has been also explained in Ref.~\cite{us1} where the contributions for the $s$-, $t$- and $u$-channels are explicitly obtained in SU(2) first. The SU(3) generalization of Eq.~(\ref{rhon}) and its application to VB systems showed that all these amplitudes make important contributions to the solution of the Bethe-Salpeter equations. We, thus, consider here the VB interaction as the sum of the amplitudes obtained from $s$-, $t$-, and $u$-channel diagrams and the contact interaction:
 \begin{equation}
 V_{VB} = V_{t} + V_{CT} + V_{s} + V_{u}.\label{vsum}
 \end{equation}
All these amplitudes are given in Ref.~\cite{us1} and thus we refer the reader to that article for more details.

Finally, we calculate the PB amplitudes using the Weinberg-Tomozawa theorem and considering the lowest order chiral Lagrangian, exactly as done in Ref.~\cite{inoue}, which leads to an amplitude of the form
\begin{equation}
V^{PB}_{ij}= -C^{PB}_{ij} \frac{1}{4 f_{i} f_{j}} ( 2\sqrt{s} - M_{i} - M_{j}) \sqrt{\frac{M_i  + E_i}{2 M_i}} \sqrt{\frac{M_j  + E_j}{2 M_j}} 
\end{equation}
where, $E_i (E_j)$ and $M_i (M_j)$ represent the energy (in the center of mass frame) and mass of the baryon in the initial (final) state. 

Although the $C^{PB}_{ij}$ coefficients for the $PB$ systems are given in Ref.~\cite{inoue} in the charge basis, we list the corresponding ones projected in the isospin 1/2 and 3/2 basis, which we use  in the present article, in Tables~\ref{PBPB1} and \ref{PBPB2}, respectively. 
\begin{table} [htbp]
\caption{ $C^{PB}_{ij}$  coefficients of the PB $\rightarrow$ PB amplitudes in the isospin 1/2 configuration.} \label{PBPB1}
\begin{ruledtabular}
\begin{tabular}{cccccc}
&$\pi N$&$\eta N$&$K \Lambda$&$K\Sigma$\\
\hline
$\pi N$&$2$&$0$&$\frac{3}{2}$&$-\frac{1}{2}$\\
$\eta N$& &$0$&$-\frac{3}{2}$&$-\frac{3}{2}$\\
$K \Lambda$&&&$0$&$0$\\
$K\Sigma$&&&&$2$\\
\end{tabular}
\end{ruledtabular}
\end{table}
With these inputs we solve the Bethe-Salpeter equation
\begin{equation}
T = V + VGT,
\end{equation}
following the  method  used in Refs.~\cite{us1,us2,us3,eovb}. In this way, following these previous works, we take care of the fact that some vector mesons have large widths
by calculating the loops for the corresponding channels by making a convolution over the varied mass of these mesons. 
\begin{table} [htbp]
\caption{ $C^{PB}_{ij}$  coefficients of the PB $\rightarrow$ PB amplitudes in the isospin 3/2 configuration.} \label{PBPB2}
\begin{ruledtabular}
\begin{tabular}{cccccc}
&$\pi N$&$K\Sigma$\\
\hline
$\pi N$& -1&-1\\
$K\Sigma$&&-1\\
\end{tabular}
\end{ruledtabular}
\end{table}

\section{Results and discussions}
With the background set up in the previous subsections, we could now start discussing the results found in our work. However, before doing that, we need to digress from this idea and discuss a little about the method followed in our work to regularize the loops in the Bethe-Salpeter equations, which are divergent in nature. One usually resorts to using a cut-off or a subtraction constant to calculate the loops in the scattering equations and these parameters are usually fixed by fitting relevant experimental data. This strategy was indeed followed  in the previous works where PB and VB systems were studied independently. Thus, in principle, we could use the parameters fixed in those works to study the coupled systems here but, as we discuss below, we come across some difficulties in doing so.

Let us start the discussion with the PB systems. In the earlier study of nonstrange PB systems based on the lowest order chiral Lagrangian~\cite{inoue}, the subtraction constants were constrained by fitting the $\pi-N$ amplitudes and as a result  a pole in the complex plane was found which was associated to the $N^*(1535)$ resonance. In a later work~\cite{hyodohosaka} it was analyzed that  the subtraction constants used in Ref.~\cite{inoue} indicate that $N^*(1535)$ does not seem to fit in the picture of a  dynamically generated resonance in PB systems. It was further investigated that the values of the subtraction constants used in Ref.~\cite{inoue} can be interpreted as adding an $s$-channel pole  to  the formalism (in other words, adding a new particle participating in the scattering), which does not originate from the PB dynamics.  This interpretation can be quickly seen if we consider a single channel two-particle scattering, in which case the Bethe-Salpeter equation can be rearranged as 
 \begin{equation}
 T=\frac{1}{V^{-1} - G}.\label{hh}
 \end{equation}
As  discussed  in Ref.~\cite{hyodohosaka}, the $G$-function, which is a real number at energies below the threshold, must be negative if we assume that there is no contribution coming from any states apart from the two scattering particles. Based on such basic principles  
of the scattering theory,  a different  scheme was proposed  in Ref.~\cite{hyodohosaka} for determining
  the subtraction constants to calculate $G$. This method ensures no contribution from the $s$-channel poles in the intermediate scattering and it does not require fitting the data. This scheme was named  in Ref.~\cite{hyodohosaka} as the ``natural renormalization scheme". 
To simplify  the further discussion, let us denote the subtraction constants and loop functions obtained in this scheme as $a^i_{\rm nat}$ and $G^i_{\rm nat}$, respectively, where $i$ symbolizes the propagating channel.  As in Ref.~\cite{hyodohosaka},  we call the subtraction constants  fixed to reproduce the data  as ``phenomenological" ones and label them as $a^i_{\rm pheno}$ and the corresponding loops as $G^i_{\rm pheno}$.

In this way,  when the $G_{\rm pheno}$-function differs from $G_{\rm nat}$ by a constant, say, $\Delta a$,  we can write $G_{\rm pheno} = G_{nat} + \Delta a$, in which case   Eq.~(\ref{hh}) becomes 
  \begin{equation}
 T= \frac{1}{V^{-1} - G_{\rm pheno}} = \frac{1}{ V^{-1} - \left( G_{nat} + \Delta a \right)}
 \end{equation}
which can be rearranged as
  \begin{equation}
 T=\frac{1}{ \left( V^{-1}  - \Delta a \right) - G_{nat}}.
 \end{equation}
 The term $\left( V^{-1}  - \Delta a \right)$ acts like a redefined kernel of the Bethe-Salpeter equation. Thus, a deviation of $G_{\rm pheno}$ from $G_{nat}$ (for $\Delta a>0$ ) can be interpreted \cite{hyodohosaka} as a modification of  the two-particle interaction. Considering the standard form of the Weinberg-Tomozawa  meson-baryon interaction: $V~\propto~(\sqrt{s}-M)/f^2$, where  $M$ and $f$ represent the mass of the baryon and the decay constant of the meson, respectively,  it can be easily shown that the modified  kernel can contain a Castillejo-Dalitz-Dyson   ($s$-channel) pole \cite{hyodohosaka}.  This seems to be precisely the case of the PB study of Ref.~\cite{inoue}. 
 
Since our purpose is to study the contribution of the vector mesons in understanding the nonstrange resonances, it would not be very useful to work in a formalism which already requires Castillejo-Dalitz-Dyson poles to explain these resonances. One alternative way would be to start by calculating the PB loops in the natural renormalization scheme of Ref.~\cite{hyodohosaka}. Let us see what results we obtain in this case.

 
Now, in case of the VB systems too, for the sake of uniformity we stick to the scheme of Ref.~\cite{hyodohosaka} instead of  using the subtraction constants of our (and other) previous works \cite{us1,eovb} on nonstrange  VB systems, although their values ($a=-2$) are not very different from the ``natural $a$" values (given in Table~\ref{anat} for both PB and VB channels). We should remind the reader that in the natural scheme of Ref.~\cite{hyodohosaka} the regularization scale, present in the loop function, is set to the mass of the baryon.

\begin{table}[h]
\caption{The ``natural" subtraction constants for PB and VB channels within the conditions explained in Ref.~\cite{hyodohosaka}. }\label{anat}
\begin{ruledtabular}
\begin{tabular}{cc||ccc}
PB Channel & Subtraction constant ($a$) &VB Channel & Subtraction constant ($a$) \\\hline
$\pi N$                & -0.3976 &     $\rho N$             & -1.5843      \\
$\eta N$		 & -1.239   &     $\omega N$       & -1.60145  \\
$K \Lambda$     & -1.143   &     $\phi N$              & -1.91566   \\
$K \Sigma$        & -1.138   &     $K^* \Lambda$ & -1.63265    \\
  			 &		 &     $K^* \Sigma$    & -1.59025    
\end{tabular}
\end{ruledtabular}
\end{table}

Let us now discuss the results we obtain by solving the Bethe-Salpeter equations with the interaction kernels of Eq.~(\ref{vsum}) and loops obtained with the subtraction constants listed in Table~\ref{anat}. 
We give the poles obtained in our study for total isospin 1/2, in two cases in Table~\ref{results_nat}: (1) when PB-VB systems are not coupled (labeled by $g_{KR}$ = 0 coupling) and (2) when they are coupled (labeled by $g_{KR}$ = 6 as given by Eq.~(\ref{gkr})).

As can be seen from Table~\ref{results_nat}, we find a wide pole around 1650 MeV in the PB channels and a very narrow pole in the VB channels in a close vicinity when the two systems are uncoupled. None of these poles can be related to known resonances.
\begin{table}[h!]
\caption{Poles and  their couplings to PB and VB channels when the Bethe-Salpeter equations are solved with loops calculated within the dimensional regularization method with the subtraction constants listed in Table~\ref{anat}.}\label{results_nat}
\begin{ruledtabular}
\begin{tabular}{ccc|cc}
PB-VB coupling&\multicolumn{2}{c|}{$g_{KR}$ = 0}&  \multicolumn{2}{c}{$g_{KR}$ = 6}\\\hline
Pole(MeV) $\longrightarrow$& $1581 -i2$ &$1649. - i130$ &$1548 - i101$&$1563 -i17$\\
\hline
Channels(Threshold) $\downarrow$&\multicolumn{4}{c}{Coupling}\\
\hline
$\pi N$(1076)               &$ 0.0 + i 0.0$   & $-1.0 + i 0.7$  &$-1.2 + i 0.2$ & $ 0.1 - i 0.4$\\ 
$\eta N$(1486)             &$ 0.0 + i 0.0$   & $-3.2 -  i 0.1$  &$-1.3 - i 3.0$  &$ 1.1 - i 0.8$\\
$K \Lambda$(1612)    &$ 0.0 + i 0.0$   & $ 1.5 + i 0.8$  &$-0.4 + i 3.6$ &$-1.0 - i 0.3$\\
$K \Sigma$(1689)       &$ 0.0 + i 0.0$   & $ 4.7 + i 0.3$ &$ 1.3 + i 1.5$ &$-0.6 + i 1.7$\\
$\rho N$(1709)            &$-0.3 - i 0.0$    & $ 0.0 + i 0.0$ &$ 0.4 + i 0.1$  &$-0.6 + i 0.2$\\
$\omega N$(1721)      &$-2.1 - i 0.0$   &$ 0.0 + i 0.0$  &$ 1.0 + i 1.0$  &$-2.5 + i 0.3$\\
$\phi N$(1959)             &$ 3.2 + i 0.0$  &$ 0.0 + i 0.0$   &$-0.7 - i 1.5$  &$ 3.7 - i 0.3$\\
$K^* \Lambda$(2008)&$ 1.4 + i 0.0$   &$ 0.0 + i 0.0$   &$-3.6 - i 2.3$  &$  1.9 - i 1.1$\\
$K^*\Sigma$(2085)    &$ 5.9 + i 0.0$   &$ 0.0 + i 0.0$   &$ 3.0 - i 2.8$  &$ 6.0 + i 0.9$\\
\end{tabular}
\end{ruledtabular}
\end{table}

When the coupling between the PB and VB channels is switched on (by allowing $g_{KR} = 6$ in Eq.~(\ref{Vpbvb})), 
these two closely spaced poles move in the complex plane to new positions: one of them ends up at $1548 - i101$ MeV and another  at $1563 -i17$ MeV. The former of these two new poles could be related to the $N^*(1535)$ but the latter one  cannot be identified with the next known nucleon resonance with spin-parity $1/2^-$, $N^*(1650)$. Moreover, we find that the known experimental data related to the $\pi N$ amplitude cannot be well reproduced although the resulting cross sections on the $\pi^- p \to \eta n$ and $\pi^- p \to K^0 \Lambda$ reactions are relatively closer to the data. The discrepancy between the experimental data and our results obtained within the natural renormalization scheme implies that some information is missing in our formalism.

The question now arises if the discrepancy between our results and the experimentally known facts can be reduced by allowing the subtraction constants to vary and if, by doing that,  an $s$-channel pole would appear in the formalism. To check this, we treat the subtraction constants for all 9 channels as free parameters which are to be fixed by requiring a fit to the experimental data. We shall later see if it is possible to make an interpretation of  those parameters (in line with Ref.~\cite{hyodohosaka}).

In order to find the new subtraction constants, we look for the best $\chi^2$ fit to  the  data set consisting of  the  isospin 1/2 and  3/2 $\pi N$ amplitudes, and the $\pi^- p \to \eta n$  and $\pi^- p \to K^0 \Lambda$  cross sections in the energy region of the low-lying resonances. In doing so, we stick to using the mass of the baryon of each channel as the regularization scale, such that we can conveniently check if we depart from the basic idea of  the natural renormalization scheme of Ref.~\cite{hyodohosaka}. To make this fit we consider PB and VB as coupled systems, i.e., we fix $g_{KR}= 6$.  In this way, we constrain the VB amplitudes too although the data set consists of reactions involving the PB channels only.  {

The best fit is obtained for the subtraction constants given in Table~\ref{afit}. 
\begin{table}[h!]
\caption{The  subtraction constants which give the best $\chi^2$ fit to  the experimental data on the  $\pi N$ amplitudes in isospin 1/2 as well as 3/2 and on the $\pi^- p \to \eta n$, $\pi^- p \to K^0 \Lambda$ reactions. The corresponding regularization scales are the baryon masses. }\label{afit}
\begin{ruledtabular}
\begin{tabular}{cc||ccc}
PB Channel & Subtraction constant ($a$) &VB Channel & Subtraction constant ($a$) \\\hline
$\pi N$                & -1.955&     $\rho N$             & -0.45      \\
$\eta N$		 & -0.777   &     $\omega N$       & -0.955  \\
$K \Lambda$     & -4.476   &     $\phi N$              & -2.972   \\
$K \Sigma$        & -1.945   &     $K^* \Lambda$ & -0.184   \\
  			 &		 &     $K^* \Sigma$    & -1.152    
\end{tabular}
\end{ruledtabular}
\end{table}
Although these values differ  from the ones given in Table~\ref{anat}, it is interesting to notice that all of them are negative numbers, which means that we get the loop functions with  negative values below the respective thresholds (at least in the neighborhood of the threshold region) as required in the scheme of Ref.~\cite{hyodohosaka}.  This finding indicates that the generation of the resonances in this formalism can be partly attributed to the meson-baryon dynamics.

Let us now discuss the results obtained with these subtraction constants. We begin by showing in Fig.~\ref{fig:tpin} the real (imaginary) parts of the $\pi N$ amplitudes,  obtained by solving the coupled channel Bethe-Salpeter equations,
\begin{figure}[h!]
\includegraphics[width= 8cm, height=6.7cm]{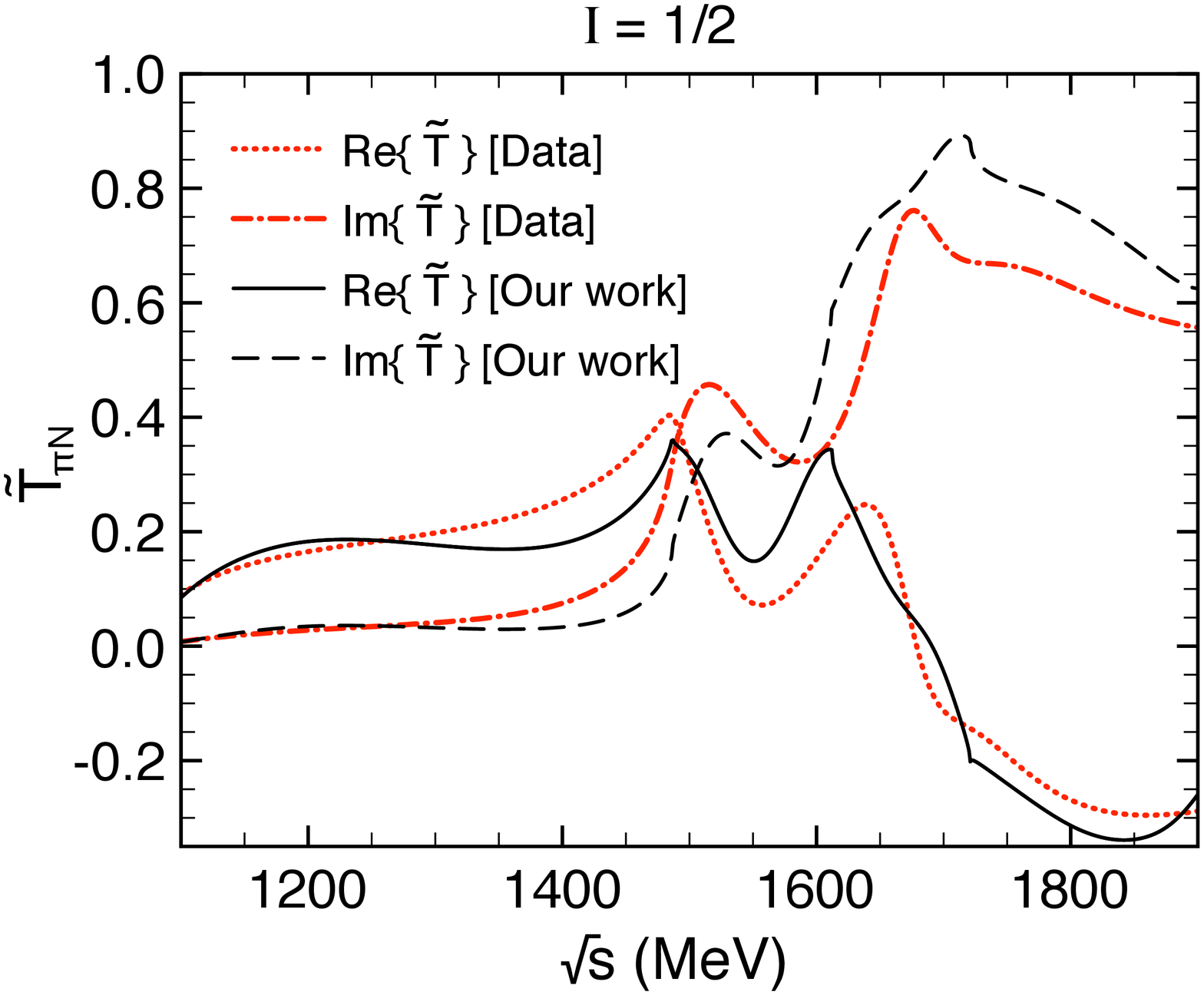}
\includegraphics[width= 8cm, height=6.7cm]{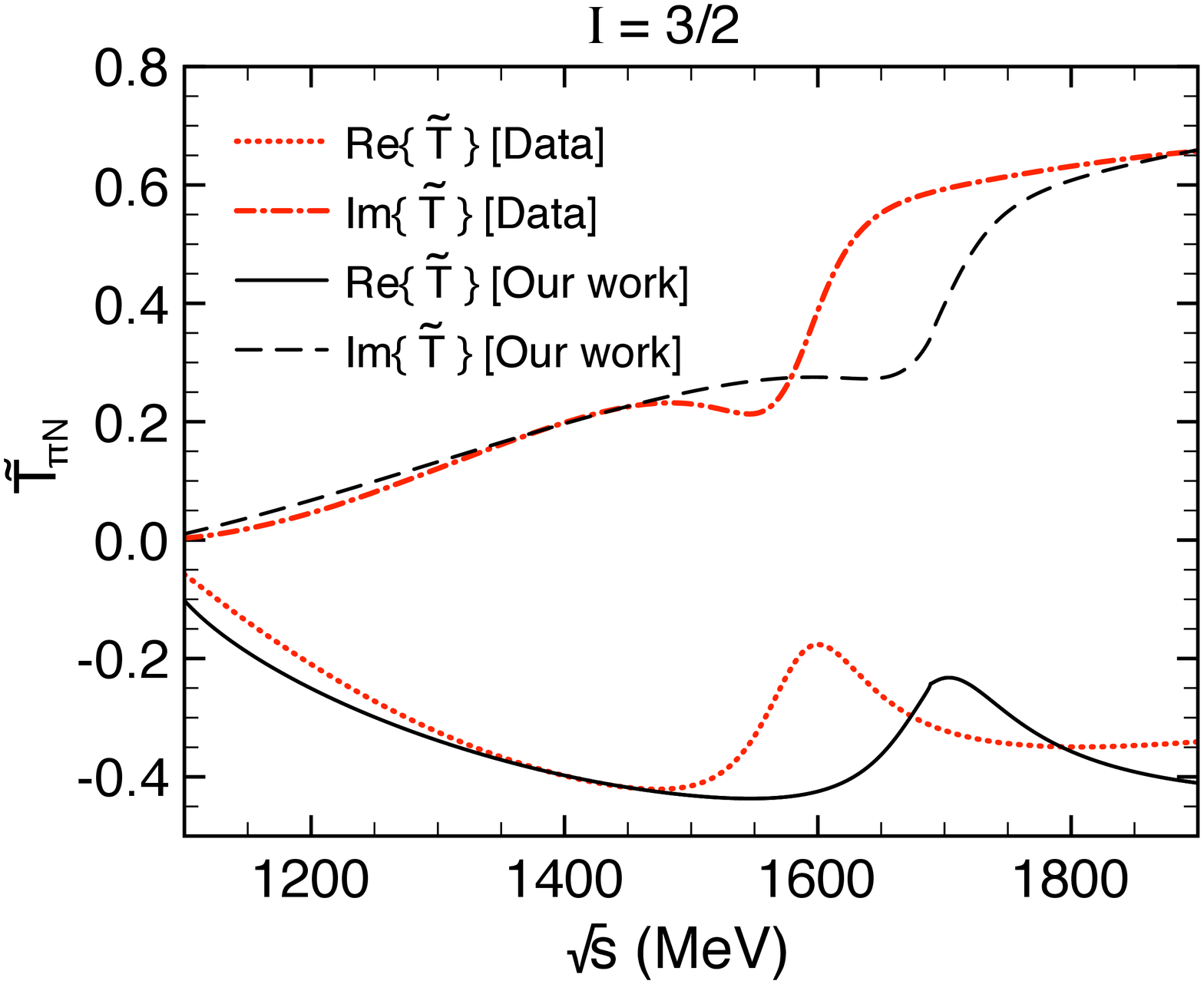}
\caption{Spin half $\pi N$ amplitudes for the isospin 1/2 (left panel) and 3/2 (right panel) configurations. The  dotted (dash-dotted) lines represent experimental data from Ref.~\cite{arndt} on the real (imaginary) part of the $\pi N$ amplitudes. The corresponding amplitudes  obtained in our work (renormalized through Eq.~(\ref{eq:renorm})), using the subtraction constants given in Table~\ref{afit}, are shown as the solid and dashed lines. }\label{fig:tpin}
\end{figure}
as solid (dashed) lines for isospin 1/2 and 3/2. Figure~\ref{fig:tpin} also shows the data \cite{arndt} on the real (imaginary) part of these amplitudes by the dotted (dash-dotted) lines.  The dimensionless amplitudes (denoted by $\tilde{T}$) shown in  Fig.~\ref{fig:tpin} are related to  the amplitudes obtained in our formalism ($T$) through
\begin{equation}
\tilde{T}_{if} (\sqrt{s}) = - T_{if} (\sqrt{s}) \sqrt{\frac{M_i q_i}{4 \pi \sqrt{s}}}  \sqrt{\frac{M_f q_f}{4 \pi \sqrt{s}}}, \label{eq:renorm}
\end{equation}
where $M_i$ ($M_f$) and $q_i$ ($q_f$) represent the mass of the baryon, and the center of mass momentum, in the initial (final) state. As can be seen from Fig.~\ref{fig:tpin},  the behavior of the $\pi N$ amplitudes gets reasonably reproduced up to about 2 GeV.  The important point to be mentioned here is that  we could not reproduce the behavior of the data beyond 1550 MeV  by considering  the PB dynamics alone (decoupled to VB systems).
This is in agreement with the previous study of PB systems involving the lowest order chiral Lagrangian  \cite{inoue}, where a reasonable fit to the  data was obtained only up to a total energy of 1550 MeV.  Our work shows  that the coupling of the vector mesons to the low-lying resonances plays an important role in obtaining  a better agreement with the experimental data. 

Next, we show the total cross sections of the $\pi^- p \to \eta n$ and $\pi^- p \to K^0 \Lambda$ reactions as a function of the beam momentum (denoted by ${\rm P}_{\rm Lab}$) in Fig.~\ref{fig:xnpb}. 
\begin{figure}[h!]
\includegraphics[width= 0.45\textwidth]{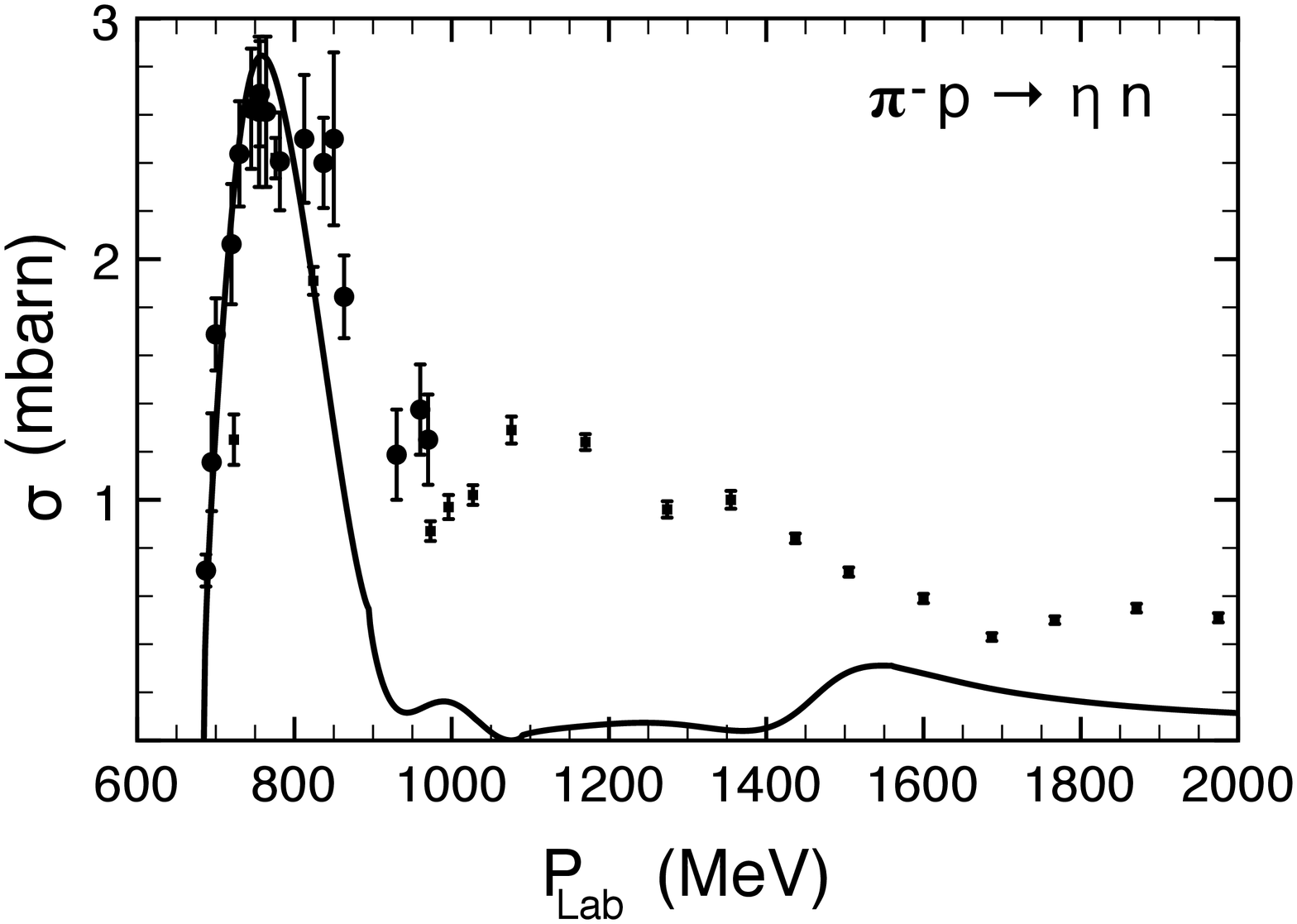}
\includegraphics[width= 0.45\textwidth]{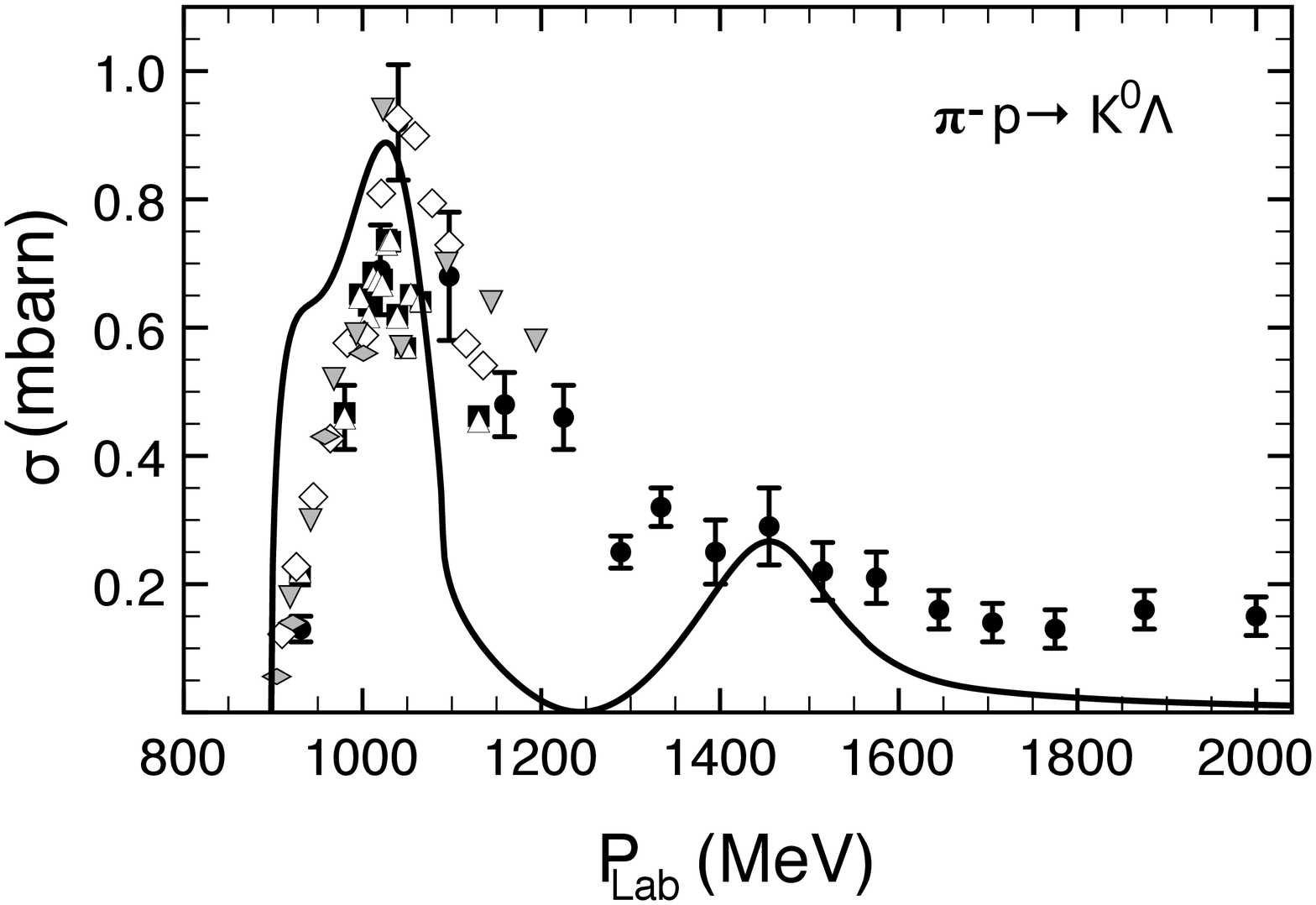}
\caption{Total cross sections for the $\pi^- p \to \eta n$ (left panel) and $\pi^- p \to K^0 \Lambda$ (right panel) reactions as a function of the beam momentum (${\rm P}_{\rm Lab}$). The experimental data have been taken from Refs.~\cite{data1,data2,data3,data4,data5}.}\label{fig:xnpb}
\end{figure}
For  the $\chi^2$-fitting we have used the experimental data on the  $\pi^- p \to \eta n$ reaction up to $\sqrt{s} \sim 1550$ MeV, which corresponds to the beam momentum (shown in Fig.~\ref{fig:xnpb}) of about 865 MeV.  It can be seen that the data near the threshold is well reproduced. In addition, a bump structure beyond that energy gets developed  around  the beam momentum  of 1600 MeV, which corresponds to a total energy of $\sim$ 1930 MeV.  We shall later discuss about the poles found in the complex  plane, which would help in understanding if this bump can be related to any of the known resonances. 

In case of  the $\pi^- p \to K^0 \Lambda$ reaction, we considered data up to  the total energy of 1760 MeV (which corresponds to the beam momentum of 1160 MeV) in the  $\chi^2$ fitting, which lead to finding the subtraction constants given in Table~\ref{afit}.  In this case too, a peak is seen beyond the threshold energy region, near the beam momentum of 1450 MeV (or $\sqrt{s} = 1900$  MeV), which seems to be in good agreement with the data. It should be mentioned in this context that in some of the recent works \cite{manley,micha} an important contribution from the  $1/2^+$ $N^*(1710)$ resonance to the $\pi^- p \to K^0 \Lambda$ reaction cross section  has been found.  However,  our formalism is restricted to s-wave meson-baryon interactions (which generates $1/2^-$ resonances). We also miss the $\pi\pi N$ system where the $N^*(1710)$ has been found to  get dynamically generated \cite{us1710}. The addition of the $\pi\pi N$ channel may help in better reproducing  the data. This possibility should be explored in future.

In order to better understand the results shown in the Figs.~\ref{fig:tpin} and \ref{fig:xnpb} we should look for poles in the complex plane. The poles obtained in our study and their couplings to the different channels are given in Table~\ref{poles},
\begin{table}[h!]
\caption{Poles and  their couplings to PB and VB channels when the Bethe-Salpeter equations are solved with loops calculated within the dimensional regularization method with the subtraction constants given in Table~\ref{afit}. The $\times$ symbols signify  no coupling of the resonance to the channel due to isospin violation. The $\#\#$ superscript indicates that the mass and the width of the state has been found from the amplitudes obtained on the real axis.}\label{poles} 
\begin{ruledtabular}
\begin{tabular}{c|r|rr|rr|c}
&\multicolumn{5}{c|}{Isospin 1/2}&Isospin 3/2\\\hline
Poles(MeV) $\longrightarrow$&$1504 - i55$ & $1668 - i28$&$1673 - i67$&$1801 - i96$&$1912 -i54$&$1689 - i56^{\#\#}$\\
Resonances associated&$N^*(1535)$&\multicolumn{2}{c|}{$N^*(1650)$}&\multicolumn{2}{c|}{$N^*(1895)$}&$\Delta(1600)$\\
\hline
Channels(Threshold) $\downarrow$&\multicolumn{5}{c}{Couplings}\\
\hline
$\pi N$(1076)               & $ 0.9 - i 0.3$   & $-0.5 - i 0.5$    & $ 1.3 - i 0.6$  & $ 0.5 + i 0.3$   & $ 0.1 - i 0.5$  &$-0.6 + i 0.3$\\ 
$\eta N$(1486)             & $ 1.3 - i 1.2$   & $-0.2 + i 1.2$   & $-1.3 - i 0.4$  & $-0.3 - i 0.2$    & $-0.2 - i 0.7$  &$    \times    $\\
$K \Lambda$(1612)    & $-0.7 - i 0.6$   & $-1.0 + i 0.7$   & $-1.2 - i 1.1$  & $-0.5 - i 0.6$   & $-0.7 + i 0.3$ &$  \times   $\\
$K \Sigma$(1689)       & $-2.5 + i 0.7$   & $-0.2 - i 0.7$   & $ 0.7 + i 0.6$  & $ 0.1 + i 0.2$   & $ 0.7 - i 0.6$  &$-0.9 + i 0.2$\\
$\rho N$(1709)             & $-1.2 - i 1.1$   & $ 3.8 + i 1.5$   & $-2.7 + i 1.9$  & $ 0.2 + i 0.5$   & $ 0.4 + i 0.2$  &$-3.0 - i 0.1$\\
$\omega N$(1721)      & $-0.8 - i 1.4$   & $-2.2 + i 1.7$   & $-2.8 - i 3.0$  & $-0.9 - i 1.3$     & $-0.5 + i 0.1$  &$  \times  $\\
$\phi N$(1959)             & $ 1.4 + i 2.2$   & $ 4.1 - i 2.7$    & $ 4.5 + i 5.2$ & $ 2.1 + i 1.8$    & $ 0.9 - i 0.2$   &$  \times $\\
$K^* \Lambda$(2008) & $ 1.1 + i 1.2$  & $ 4.1 - i 2.6$    & $ 4.5 + i 4.0$ & $ 2.5 + i 3.0$    & $ 4.0 + i 0.4$  &$  \times $\\
$K^*\Sigma$(2085)     & $-1.2 + i 0.6$   & $ 2.3 - i 2.8$   & $ 4.9 + i 1.8$ & $ 4.6 + i 0.3$    & $-3.0 - i 1.0$   &$~3.3 + i 0.0$\\
\end{tabular}
\end{ruledtabular}
\end{table}
 which also shows the known resonances to which these poles can be related. We find evidence for three isospin 1/2 and one isospin 3/2 resonance. Let us discuss how the  properties of the poles found in the present work compare with those of the  known resonances.  
\begin{enumerate}[(i)]
\item{\bf \underline{Isospin 1/2}}
\begin{enumerate}[(1)]
\item{Let us begin with the pole found at  $1504 - i55$ MeV.   The properties of this state are in good agreement with those of  the negative parity $S_{11}$ $N^*(1535)$ resonance  \cite{pdg}, which is known to correspond to a pole with the mass between 1490  to 1530 MeV and full width between 90 to 250 MeV \cite{pdg}. Also, the experimental studies find that this resonance has similar branching ratios to the channels $\pi N$ and $\eta N$ \cite{pdg}. This is compatible with our finding of  the coupling of $\pi N$ to this state, whose strength is about half of  the  one obtained for the $\eta N$ channel.  In fact, we calculate the branching ratios of our state to these decay channels by using the imaginary part of the corresponding amplitude in order to take the width of the resonance into account \cite{ollernpa}. As a result we obtain a branching ratio of 43$\%$  for the $\pi N$  and 55$\%$ for the $\eta N$ channel. These results are in excellent agreement with those given in Ref.~\cite
{pdg}.}

\item{Next we find a twin pole with positions: $1668 - i28$ MeV and $1673 - i67$ MeV (see Fig.~\ref{twonstars}).  We find that the appearance of a twin pole is unavoidable in this energy region while minimizing the $\chi^2$.  We relate these states to the next low-lying $S_{11}$  resonance, $N^*(1650)$, which, according to Ref.~\cite{pdg}, corresponds to a pole with the real part  ranging between 1640-1670 MeV and the full width varying from 100 to 170 MeV.  In fact, as pointed out in Ref.~\cite{pdg}, the two pole nature of this resonance has earlier  been discussed in Ref.~\cite{arndt}.  Our results are similar to the poles found for this resonance in Ref.~\cite{arndt}: $1673 -i41$, $1689 - i96$ MeV. 
\begin{figure}[h!]
\includegraphics[width= \textwidth]{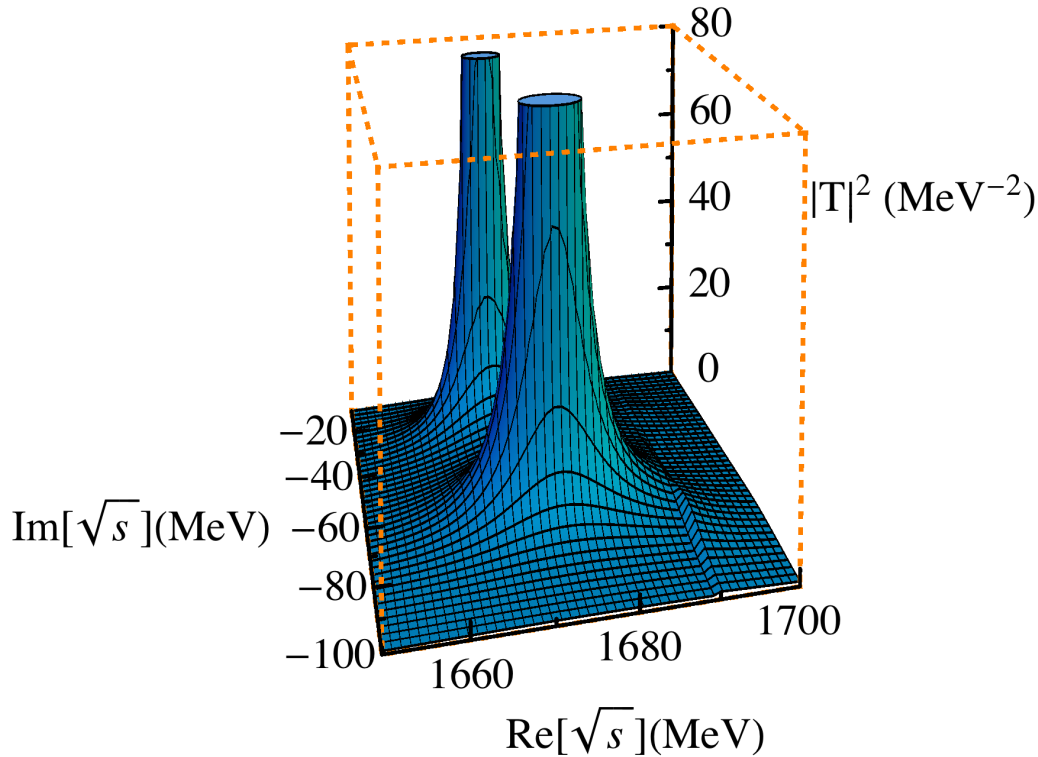}
\caption{Double pole structure related to $N^*(1650)$.}\label{twonstars}
\end{figure}

It should be emphasized here that a dynamically generated nature for  both $N^*(1535)$ and $N^*(1650)$ 
has been found within a formalism based on the lowest order chiral Lagrangian which requires fixing of minimum number of parameters. We would like to mention again here that a good fit beyond 1600 MeV  is not found when considering PB channels only (which is similar to the findings of Ref.~\cite{inoue}). In this sense, we can consider the finding of poles corresponding to $N^*(1650)$, together with a reasonable fit obtained to the data on $\pi N$ amplitudes beyond 1600 MeV, as a success of our model.
Although, as can be seen from Figs.~\ref{fig:tpin} and \ref{fig:xnpb}, our amplitudes are not in perfect agreement with the experimental data, and the subtraction constants used here (Table~\ref{afit}) differ from the natural $a$ given in Table~\ref{anat}. This implies that these resonances cannot be interpreted as purely dynamically generated ones. There is some information missing in our formalism.  However, our findings do indicate that adding the vector mesons to the coupled channel space improves the compatibility with the experimental data.}

\item{Further, we find a resonance at $1801 - i96$ MeV and another one at  $1912 -i54$ MeV.  Little is known about $1/2^-$ nucleon resonances beyond $N^*(1650)$. This becomes evident if one looks at the  note in Ref.~\cite{pdg} under the next, and the only other, $1/2^-$ nucleon resonance, $N^*(1895)$, which says that any structure in the $S_{11}$~wave found above 1800 MeV is listed together. 
Although we find two resonance poles beyond 1800 MeV, it might be difficult to identify them in the experimental data due to their large overlapping widths and a single peak might be seen as a result of the interference of the two.
In fact, our results on the $\pi^- p \to K^0 \Lambda$ reaction show only one resonance peak around 1900 MeV (which corresponds to
the beam momentum of about 1400 MeV) in the total cross section plot shown in the right
panel of Fig. 2. As can be seen, the cross section near the peak position  is compatible with the data.
Also a bump is seen in the  $\pi^- p \to \eta N$  cross sections around this energy.  
We have also calculated the total cross sections for the  $\pi^- p \to  \omega n$ and $\pi^- p \to \phi n$ reactions to check the contribution of the resonances at  $1801 - i96$ MeV and  $1912 -i54$ MeV found in our work. The results on the $\phi n$ production, to which only the tail of these resonances should contribute, are in agreement with the data (see Fig.~\ref{fig:xnvb}). 
However the data on the  $\pi^- p \to  \omega n$ reaction is not well reproduced by our results but we are here studying only $1/2^-$ resonances. It is known that several  resonances with different spin-parities contribute to the omega production \cite{shkylar,moselpenner}.
\begin{figure}[h!]
\includegraphics[width= 0.45\textwidth]{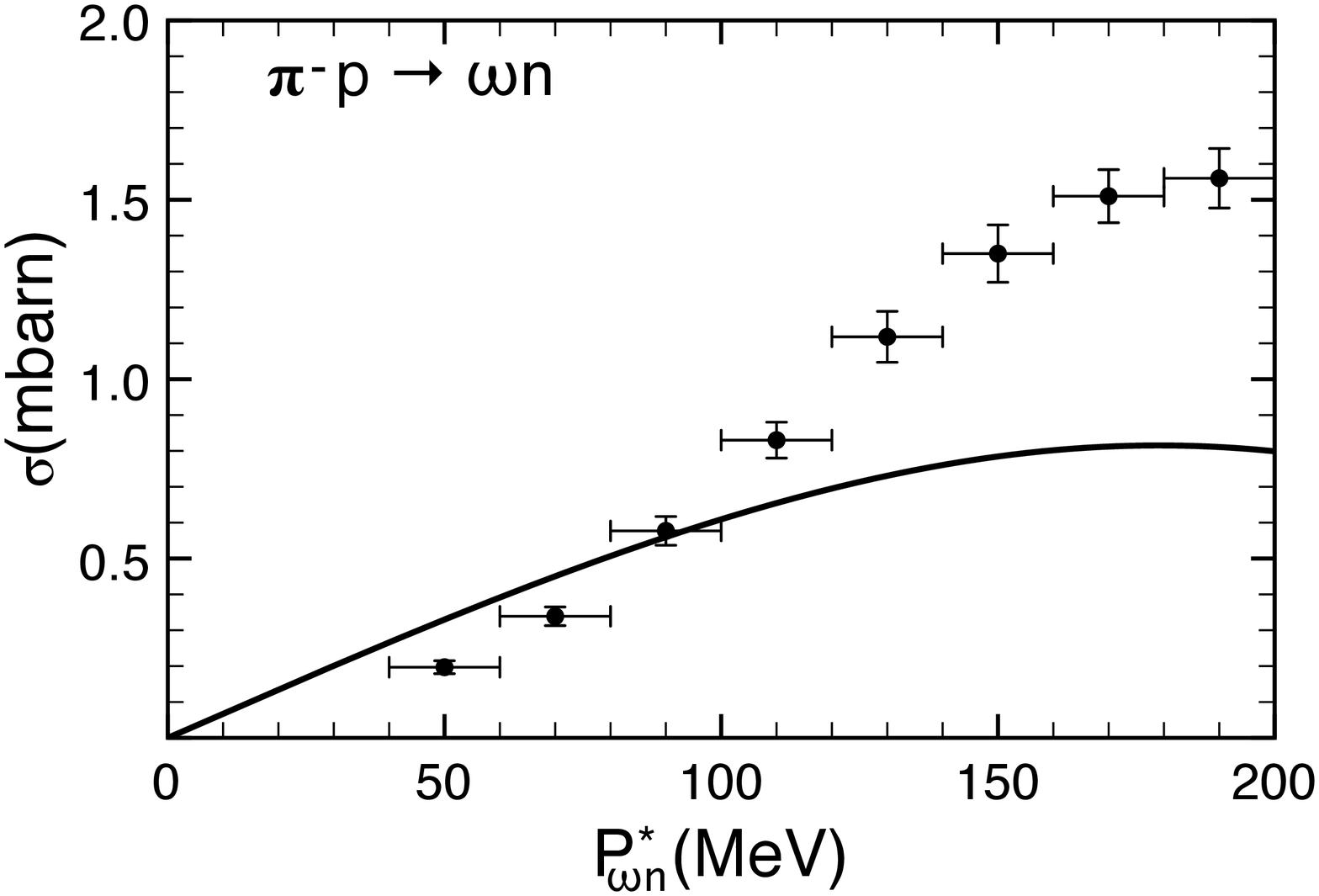}
\includegraphics[width= 0.45\textwidth]{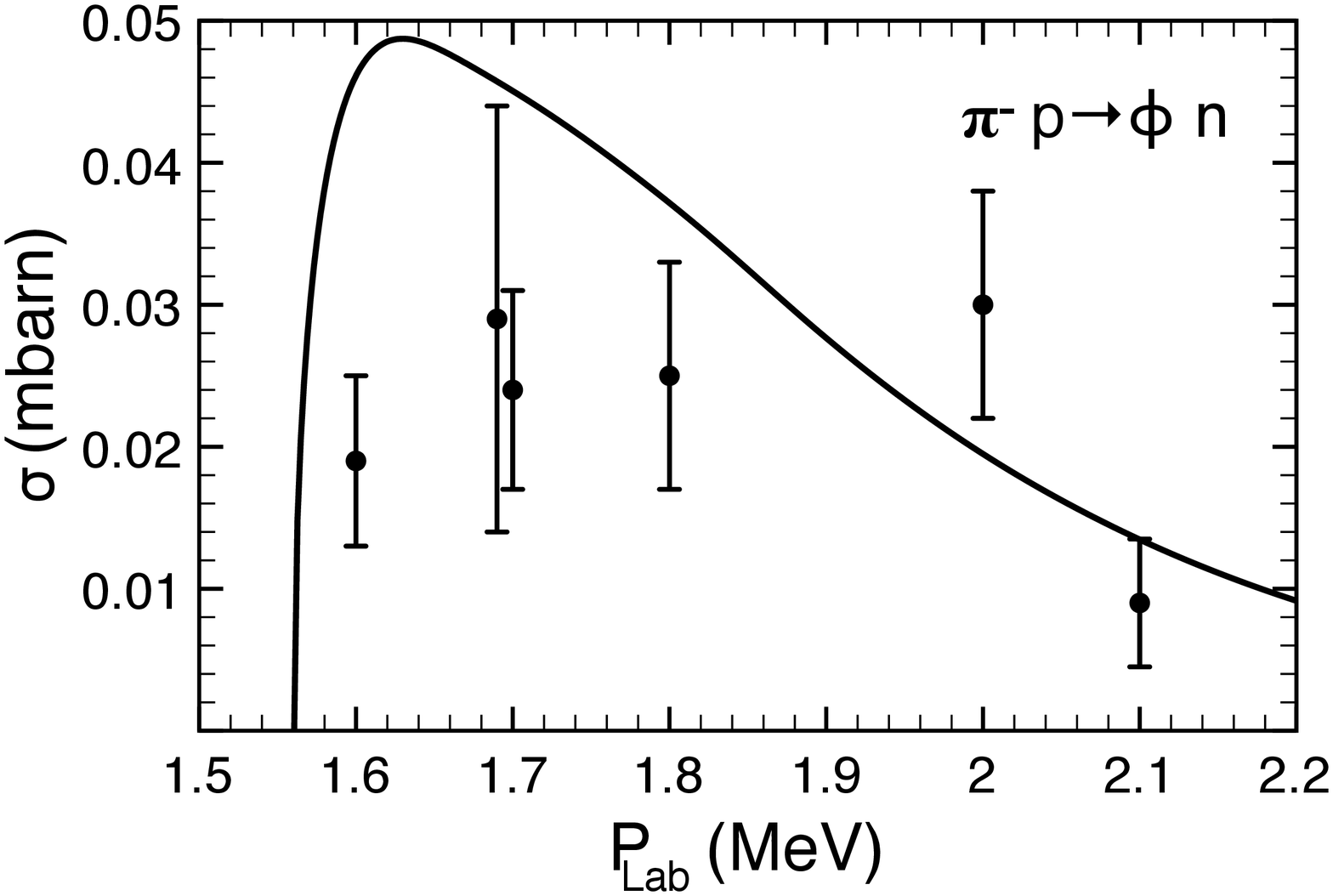}
\caption{Total cross section on the pion induced omega (shown in left panel) and phi (shown in right panel) production on a nucleon. The left panel shows the cross~section as a function of the center of mass momentum in the final state while on the right side we show the cross section as a function of the beam momentum. This is done to show our results in comparison with the experimental data as available from Refs.~\cite{data6,data7}. }\label{fig:xnvb}
\end{figure}}
\end{enumerate}

\item{\bf \underline{Isospin 3/2}}

Finally, let us discuss a bit on the pole obtained at $1689 - i56$ MeV in the isospin 3/2 configuration. The mass and width of this state has been extracted using the amplitudes obtained on the real axis. The corresponding pole in the complex plane lies very close to the threshold of the $\rho N$ channel and since we need to calculate the loop-function for this channel by convoluting over the width of the $\rho$-meson, it  gets complicated to make a reliable pole analysis in the complex plane.  This pole can be associated with the well known $\Delta (1620)$ resonance, for which the mass and the width are given in the range of 1600-1660 MeV and 130-150 MeV, respectively \cite{pdg}. The range for the pole position listed in Ref.~\cite{pdg} is (1590-1610)$ - i $(120-140) MeV.  Although our state has a slightly higher mass. We should notice that the $\Delta (1620)$  resonance is known to decay with a large branching ratio to the $\Delta \pi$ channel which is not included in our present formalism. However, a significant branching ratio for $\Delta (1620)$ has also been found to the $\rho N$ decay channel, which is compatible with the large coupling shown in Table~\ref{poles}. It is also important to notice that our work shows that a $S_{31}$ resonance with negative parity,  and mass near 1690 MeV, couples  relatively weakly to the PB channels. This is also in agreement with the small branching ratio to the $\pi N$ channel \cite{pdg,jujun} in spite of the presence of a large phase space. 
\end{enumerate}

 We have also calculated the scattering lengths for the different channels, which are summarized in Table~\ref{length_1h} for the ispospin 1/2 configuration and in Table~\ref{length_3h} 

  \begin{table}[h!]
\caption{Isospin 1/2 scattering lengths obtained in the present work for different PB,VB channels. }\label{length_1h} 
\begin{tabular}{cccccccccc}\hline\hline
   PB channels               &   $\pi N$ & $\eta N$ & $K \Lambda$ &  $K \Sigma$  &   \\ 
$a^{I=1/2}_{i}$ (fm)  &   $0.22$ & $0.4 + i 0.2$ & $-0.05 + i 0.26$ &  $-0.43 + i 0.05$  &    \\\hline
   VB channels               &      $\rho N$ &  $\omega N$ &  $\phi N$ &  $K^* \Lambda$  &  $K^*\Sigma$\\ 
$a^{I=1/2}_{i}$ (fm)  &       $-0.62 +  i1.2$ &  $-0.61  + i0.37$ &  $-0.22 + i0.09$ &  $-0.66 + i0.4$  &  $-0.56 + i0.2$\\
\hline\hline
\end{tabular}
\end{table}
\begin{table}[h!]
\caption{Isospin 3/2 scattering lengths obtained in the present work for different PB,VB channels. }\label{length_3h} 
\begin{tabular}{cccccccccc}\hline\hline
                   &   $\pi N$  &  $K \Sigma$  &    $\rho N$ &  $K^*\Sigma$\\ \hline
$a^{I=3/2}_{i}$ (fm)  &   $-0.27$ & $-0.1 + i 0.08$ & $0.4 + i 1.13$ &  $-0.32 +i 0.13$  \\
\hline\hline
\end{tabular}
\end{table}
for the isospin 3/2 case. 
As can be seen from these tables, we obtain $a_{\eta N} = 0.4 + i 0.2$ fm, $a^{I=1/2}_{\pi N} = 0.22$ fm  which are in reasonable agreement with known values. The 
  $\eta N$ scattering length is known to vary in a wide range:   $a_{\eta N; {\rm known}} = (0.18 + i 0.16)$ fm to $(1.03 + i 0.49)$ fm (for a recent review on  $\eta$-nuclear interaction see Ref.~\cite{scatlen1}). The isospin 1/2 $\pi N$ scattering length is known from the experimental study \cite{scatlen2} to be  $a^{I=1/2}_{\pi N; {\rm exp}} = 0.25 \pm 0.005$ fm.  On the other hand, the $\pi N$ scattering length found here, in the  isospin 3/2 case, $a^{I=3/2}_{\pi N} = - 0.27$ fm, is higher as compared to the experimentally known value $a^{I=3/2}_{\pi N; {\rm exp}} = -0.132 \pm 0.0132$ fm \cite{scatlen2}. As mentioned earlier, the discrepancy between our results and the data indicates that the the current formalism lacks some ingredient.  
  
One  possible extension of the present formalism could be the inclusion of the decuplet baryons. An elaborate work on meson-baryon systems with strangeness $0, -1, -2, -3$, involving pseudoscalar and vector mesons together with octet and decuplet baryons, has been done in Ref.~\cite{carmen2} where dynamical generation of several $1/2^-$ resonances, including $N^*(1535)$, $N^*(1650)$, has been reported. While $s$- and $u$- channel interactions were not considered  in Ref.~\cite{carmen2} 
the authors have extended the Weinberg-Tomozawa interaction to include both vector mesons
and decuplet baryons by assuming SU(6) symmetry.  Such a symmetry consideration is useful, and the results  found in Ref.~\cite{carmen2}
should serve as a point of reference  to make comparisons in future.  It would be, thus, interesting to extend our present formalism  by  including deculplet baryons, like in Refs.~\cite{carmen1,carmen2}, in future.  
 
Before ending this section, we would like to mention a limitation of the present formalism. This is related to the  energy range in which the present formalism is reliable. We find that the isospin 1/2 $T$-matrices obtained in this work are not reliable when going far below the threshold region. There, peaks corresponding to unphysical states might be found although the interaction potential is repulsive in nature. Of course, our amplitudes can also not bring reliable information beyond 2.2 GeV, i.~e., while going far from the threshold of the heaviest VB channel, since
only s-wave meson baryon interaction is considered here, which is suitable to study the dynamical generation of baryon resonances in these systems.

\section{Summary}
We have studied meson-baryon systems as coupled channels to investigate the dynamical generation of resonances. The systems under consideration have total isospin 1/2, 3/2 and spin 1/2. The mesons and baryons interact in s-wave, which implies that the possible resonances generated in the system can have spin-parity $1/2^-$. 
The formalism consists of solving Bethe-Salpeter equations, for which the interaction kernels are obtained from the Lagrangians based on the chiral and hidden local symmetries. In order to calculate the divergent loop functions, we use the dimensional regularization scheme.  We first attempt to strictly follow the natural renormalization scheme of Ref.~\cite{hyodohosaka} to get the subtraction constants needed  to compute the loop functions. The advantage of this scheme lies in the  possibility of getting contributions from  the dynamically generated resonances exclusively.  As a result, we find two poles near 1550 MeV which cannot be related to known resonances.  Further, we cannot reproduce the available experimental data with the corresponding amplitudes.

Next, we obtain the  subtraction constants by  making a $\chi^2$-fit to the available experimental data on the $\pi N$ amplitudes (till $\sim$ 2 GeV) and on the reactions: $\pi^- p \to \eta n$ (up to 1550 MeV) and $\pi^- p \to K^0 \Lambda$ (till 1760 MeV). Although the subtraction constants found in this way differ to the ones within the natural renormalization scheme, we find that their values (and hence the loop functions) are such that the states obtained in this work can be interpreted as partly dynamically generated ones.

Consequently, we find poles in the complex plane whose properties are in good agreement  with those of some known resonances. To be specific, we find evidence for  the $N^*(1535)$ and $N^*(1650)$ resonances, with a double pole associated to the latter one.  Since the information on all the $1/2^-$ states with mass beyond 1800 MeV is put together under the resonance $N^*(1895)$ in Ref.~\cite{pdg}, it appears that there is only one known  $S_{11}$ resonance beyond 1800 MeV.  Our work provides evidence for the existence of  two $N^*$'s, one with mass 1800 MeV and other with 1912 MeV, though the large overlapping widths found for these two  resonances  show that it could be difficult to identify two distinct states in the cross sections in the 1800-1900 MeV region. 

Finally, we find a resonance with isospin 3/2 at $1689 - i56$ MeV. This state can be related to $\Delta(1620)$.

We can conclude this work by answering the question raised in the beginning of this article: does vector meson-baryon dynamics bring new information in understanding the nature of the low-lying nonstrange resonances, like $N^*(1535), N^*(1650)$. Our work indicates that with the addition of vector mesons to build the coupled channels we seem to move in the direction of understanding the low-lying $N^*$ and $\Delta$ resonances as dynamically generated states, at least partly. The next question  which might be raised now is which information can help in getting a clearer picture about the origin of the low lying nonstrange resonances.  The answer can be adding the decuplet baryons to the formalism since some resonances decay to meson and decuplet baryon channels with significant branching rations. The results found here should serve as a motivation for such an extension of the formalism in future works.

\section{Acknowledgements}
 A.~M.~T  and K.~P.~K  thank the Brazilian funding agencies FAPESP and CNPq for the financial support.
This work is partly  supported  by the Grant-in-Aid for Scientific Research on Priority Areas titled ÒElucidation of New Hadrons with
a Variety of Flavors" (E01: 21105006 for  A.~H) and (24105707 for H.~N) and the authors acknowledge the same.
A.~M.~T  and K.~P.~K  are also grateful for the partial financial support provided by the same project mentioned for H.~N above to stay in Japan, where a part of this work was accomplished.

\end{document}